\documentclass[lettersize,journal]{IEEEtran}
\usepackage{amsmath,amsfonts}
\usepackage{algorithmic}
\usepackage{algorithm}
\usepackage{array}
\usepackage[caption=false,font=normalsize,labelfont=sf,textfont=sf]{subfig}
\usepackage{textcomp}
\usepackage{stfloats}
\usepackage{url}
\usepackage{verbatim}
\usepackage{graphicx}
\usepackage{cite}
\usepackage{makecell}
\usepackage{amsmath}
\usepackage{pifont}
\usepackage{multirow}
\begin{document}

\title{CAT: Customized Transformer Accelerator Framework on Versal ACAP}

\author{Wenbo Zhang,Yiqi Liu,Zhenshan Bao
\thanks{We thank the Versal VCK 5000 board donated by the Xilinx University Program (XUP). We thank UCLA’s AMD/Xilinx heterogeneous accelerated computing cluster.}
\thanks{Manuscript received April 19, 2024; revised August 16, 2024.}}

\markboth{Journal of \LaTeX\ Class Files,~Vol.~14, No.~8, August~2024}%
{Shell \MakeLowercase{\textit{et al.}}: A Sample Article Using IEEEtran.cls for IEEE Journals}

\IEEEpubid{0000--0000/00\$00.00~\copyright~2024 IEEE}

\maketitle

\begin{abstract}
Transformer uses GPU as the initial design platform, but GPU can only perform limited hardware customization. Although FPGA has strong customization ability, the design solution space is huge and the design difficulty is high. Versal ACAP is a heterogeneous computing architecture with AI Engine as the core. It is far more flexible than GPU in hardware customization, and has better and smaller design solution space than traditional FPGA. Therefore, this paper proposes the Customized Transformer Accelerator Framework(CAT), through the CAT framework, a customized Transformer accelerator family can be derived on Versal ACAP, CAT framework has an abstract accelerator architecture design idea, which deconstructs and efficiently maps the Transformer into the hardware, which contains a variety of customizable properties. Through the customization and optimization strategy of the CAT framework, the underlying hardware and the upper model jointly constrain and decide on these customizable properties, and finally form a customized accelerator. We use a 7 nm AMD Versal ACAP VCK5000 development board to implement accelerators for different Transformer models based on the CAT framework. Experiments show that we achieve the highest throughput gains of 2.41×, 49.50×, and 1.32× compared to 8 nm Nvidia GPU A10G, 16 nm AMD FPGA ZCU102, and 7 nm AMD Versal ACAP VC190(SOTA). The highest energy efficiency gains are 7.80×, 6.19× and 1.15×, respectively.
\end{abstract}


\section{Introduction}

\IEEEPARstart{T}{ransformer}\cite{Transformer} have shown great success in NLP, CV, and LLM, but have been criticized for consuming too much computing power.  The intensive computation and huge memory overhead have posed great challenges to hardware, which has aroused a surge of research on Transformer accelerators.

Transformers were first introduced on the GPU platform, and different Transformer families(cover different area, from NLP\cite{BERT,ROBERTA} to CV\cite{VIT1,VIT2}, from deep learning networks to LLM) were derived according to different application requirements. The difficulty of accelerator design on GPU platform is mainly to solve the pressure of operator fusion and memory access when CUDA core is executed in parallel\cite{cuda}. Among them, the Flash Attention series\cite{flashattention,flashattention2} proposed by Tri Dao has become the main method in the current industry. Therefore, its high computing power cost has been criticized by the industry. FPGA and ASIC have strong hardware customization capabilities, and subsequent researchers hope to reduce the cost of computing power through system customization. Although the results are remarkable, the high degree of customization leads to insufficient flexibility and is not convenient to transfer it to family members. Correspondingly, it brings higher design cost and design difficulty. It is difficult to balance between the efficiency and flexibility in design process.

\IEEEpubidadjcol

Versal ACAP\cite{versal_acap1,versal_acap2} is a heterogeneous computing architecture with AI Engine(AIE)\cite{aie} as the core, which efficiently integrates Processing System(PS), Programmable Logic(PL) and AIE. Versal ACAP architecture is a good compromise between the GPU and FPGA. On the one hand, it provides highly customizable ability, which can be designed at the hardware level, which is not possible with GPU. At the same time, due to the support of AIE, its design granularity is coarser and better than that of FPGA and ASIC, and the design solution space is optimized. Although there has been a few works on ACAP\cite{CHARM,WideSA,SSR,EA4RCA}, it basically focuses on the development of classical high-performance computing operators, and has not designed a general architecture.

To this end, we propose the CAT framework, which can derive custom Transformer accelerator family on Versal ACAP and thus adapt to different Transformer models.We evaluate the performance of CAT on Versal ACAP. Compared with existing Transformer accelerators, CAT achieves faster inference speed and lower power consumption. In our experiments, accelerators implemented on the CAT framework were compared with 8 nm Nvidia GPU A10G, 16 nm AMD FPGA ZCU102, and 7 nm AMD Versal ACAP VC190(SOTA). We achieve the highest throughput gains of 2.41×, 49.50×, and 1.32×. The highest energy efficiency gains are 7.80×, 6.19× and 1.15×, respectively.

The main contributions of this work can be summarized as follows:

\begin{itemize}
\item{\textbf{Propose a Transformer abstract accelerator architecture for Versal ACAP.} The CAT framework has an abstract architecture that disassembles and efficiently maps transformers into hardware, and adopts a module organization scheme with customizable parallel patterns. Through this architecture, a customized Transformer accelerator family can be derived on Versal ACAP.}
\item{\textbf{Designing a top-down framework customization strategy starting from the Transformer model.} The customization strategy of CAT framework is used to customize the abstract architecture by analyzing the characteristics of the underlying hardware and the upper model, so that the final accelerator and the model show a strong pertinence and fit.}
\item{\textbf{Analyze the inference performance and energy efficiency of accelerators generated by the CAT framework.} We conduct detailed performance evaluation and analysis of different Transformer accelerators customized based on the CAT framework, and obtain better overall performance and energy efficiency compared with existing methods and hardware platforms.}
\end{itemize}

The rest of this paper is organized as follows. Section II briefly reviews related work and research background. In Chapter III, the abstract architecture design of CAT framework is given. Chapter IV introduces the customization strategy of CAT framework. We present and evaluate the experimental results in Section V. Finally, we conclude the paper in Section VI.

\section{Observation and Motivation}
\subsection{Related Work on Transformer Accelerator}

GPU is currently the mainstream platform for accelerating Transformer\cite{GPU_ACC_A10G,flashattention,flashattention2,flashattention3,MegatronLM}, Since the hardware of the GPU cannot be customized, the design of the accelerator on it is software-oriented. FlashAttention\cite{flashattention} reorders attention computation, solve the memory bottleneck of GPU when computing attention. After that, the Tri Dao team released Flash Attention-2\cite{flashattention2} in 2023, which achieved the efficiency of running GEMM operator on A100. At the same time, the softmax operator was greatly optimized to effectively remove line coupling and achieve efficient block calculation.Flash Attention-3\cite{flashattention3} has been released in 2024 to further improve efficiency.

There are a wide variety of accelerators in the field of FPGA\cite{FPGA_ACC_AUTOVITACC,FPGA_ACC_FTRANS,FPGA_ACC_HPTA,FPGA_ACC_MEVIT,FPGA_ACC_MULTCIM,FPGA_ACC_ULSEQTA,FPGA_ACC_VIA}, among which ME-ViT\cite{FPGA_ACC_MEVIT} employs a single-load strategy where the model parameters are loaded only once, intermediate results are stored on the chip, and all operations are implemented in a single processing element. Softmax and LayerNorm functions are also integrated into ME-PE to reduce the pauses between matrix multiplications. ViA\cite{FPGA_ACC_VIA} implements a novel Vision Transformer accelerator architecture on Xilinx Alveo U50 FPGA. FTRANS\cite{FPGA_ACC_FTRANS} employs an enhanced Block Circulant Matrix (BCM)- weight representation based approach that enables model compression on large-scale language representations at the algorithmic level. HPTA\cite{FPGA_ACC_HPTA} High-performance accelerator for Transformer implementation on FPGA. Analyze the structural characteristics of the network and design accelerators with configurable processing elements, optimized data selection and arrangement, and efficient memory subsystems to support various Transformer models.

Focusing on the problems existing in GPU and FPGA, AMD/Xilinx proposed Versal ACAP\cite{versal_acap1,versal_acap2}. In the field of Versal ACAP, there have been studies based on Versal ACAP\cite{CHARM,WideSA,SSR,EA4RCA,acap_impl1,acap_impl2,acap_impl3,acap_impl4,vitis_lib}, the advantages of this architecture are recognized, but there is only a few of relevant research. CHARM(FPGA '23)\cite{CHARM} by Jason Cong's team is the first work in the field of Transformer acceleration based on Versal ACAP, and SSR(FPGA' 24)\cite{SSR} is its follow-up work. CHARM effectively implements Dense Matrix Multiplication (MM) in VCK190, and implements an automatic generation mechanism to accelerate the reasoning of the upper model by calling MM operator multiple times. However, this method is often inefficient, and the communication overhead and power waste caused by multiple calls to the operator are very obvious. SSR builds the accelerator by constructing multiple computing units of the same architecture and using spatial sequential hybrid scheduling of computing units at the top level, which is the current level of SOTA. However, this method is relatively general and its effective utilization of AIE is low.

Therefore, in the following we give several observations in the field of Versal ACAP accelerated transformers.

\subsection{Observations}

\subsubsection{Transformer Implementation on Versal ACAP}
\ 
\newline
\indent 
\textbf{Observation 1}: Versal ACAP is suitable for AIE as the core of the design method, while the PL hardware adjacent to AIE as a fixed pipeline module, the organization and customization of the module as a unit can help to develop the hardware performance.

\begin{figure}[!t]
\centering
\includegraphics[width=2.5in]{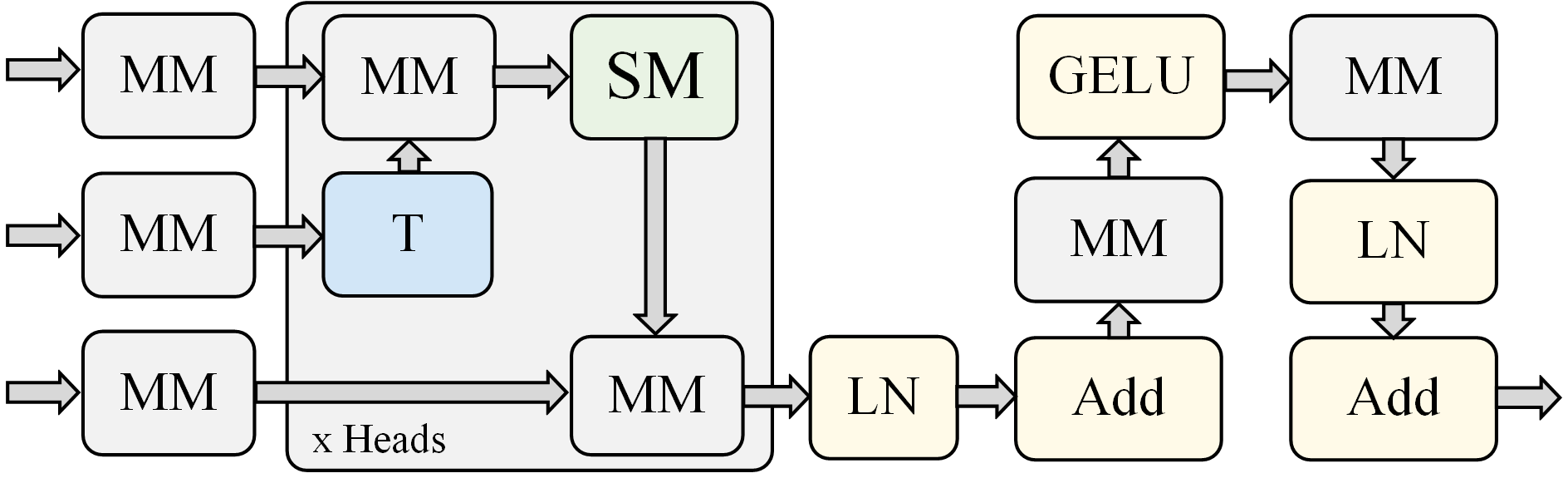}
\caption{Transformer model architecture.}
\label{fig_1}
\end{figure}

Figure 1 shows the classical Transformer model architecture. If all its operations are fused into a processing element, it will contain a large number of operators with different types and scales, and contain a large number of data flows.

The memory bandwidth of AIE (256bit/s) is less than the computational efficiency of AIE (1024bit/s), so it is more suitable for the computationally intensive matrix multiplication (MM) operation, which assumes a large part of the computational load. However, memory constrained operators such as SoftMax, LayerNorm, and GELU are more suitable to be executed on PL due to their low computation and frequent memory access.

Secondly, we find that splitting computation and communication, adopting a regular communication design pattern to reduce the number of computations interrupted, and making full use of AIE DMA(15.6TB/s) to transfer data to AIE's on-chip memory at one time can effectively increase the proportion of computation time during AIE runtime, thereby improving the performance. This was demonstrated in our previous work EA4RCA framework\cite{EA4RCA}.

At the same time, the AIE and the PL module adjacent to AIE are treated as fixed and pipelined modules, which helps to improve the efficiency of AIE in calculating MM. There are three stages in the running of AIE: sending, computing and receiving. If the three stages are organized in serial at the PL side, we can get a time of (1.1x, baseline), if they are organized in parallel at the PL side, then the time is (0.71x, 1.41 speed up), if there are more PL modules near AIE, we recommend pipelining them all. If multiple such modules exist, AIE is guaranteed to be efficient at runtime even if they are organized serially.

\textbf{Conclusion 1}: It is necessary to design various hardware efficiently by combining the architectural characteristics of Versal ACAP.

\subsubsection{Composite Application Design}
\ 
\newline
\indent 
\textbf{Observation 2}: Existing works have fully demonstrated the performance of ACAP and proposed optimization schemes, but most of them focus on a single application load. How to efficiently utilize AIE in composite applications such as transformers still needs to be explored.

In EA4RCA framework\cite{EA4RCA}, the deployment and optimization of a single application on Versal ACAP are discussed in detail, and the results of SOTA are obtained, such as MM, FFT, Filter, etc. The hardware of Versal ACAP has a single function in these applications. However, different from EA4RCA, Transformer contains many kinds of operators, even the same kind of operators with different scales, so we need to consider how to make AIE or PL bear more kinds of operators without loss of performance.

Our load analysis of Transformer shows that the number of calls of matrix multiplication operators can reach half of the total number, and the computational load occupied by matrix multiplication accounts for more than 90\% of the total. Therefore, we take the data flow path of matrix multiplication as the backbone data flow of the accelerator, and insert the other nonlinear operator modules as branches into the backbone data flow. At the same time, due to the pipelined hardware design, the insertion of these modules will not affect the overall delay, but will only increase the depth of the pipeline. Therefore, at the AIE level, the application undertaken by AIE is single to maximize the computational performance of AIE, and PL assumes a variety of other nonlinear operators of Transformer and is inserted into the data flow of matrix multiplication as branches.

\textbf{Conclusion 2}: The design of composite applications still needs to be further explored.

\subsection{Motivation}

In order to achieve higher total throughput and maintain comparable energy efficiency compared with FPGA, we choose VCK5000 board based on AMD/Xilinx Versal ACAP architecture\cite{versal_acap1,versal_acap2} as the hardware platform design accelerator. The peak throughput of VCK5000 board is on the same order of magnitude as that of GPU. In terms of power consumption, it is similar to FPGA. Versal ACAP architecture with AI Engine(AIE)\cite{aie} as the core, it efficiently integrates Processing System(PS), Programmable Logic(PL) and AIE, AIE internal core can be customized design and free combination. At the same time, it is equipped with flexible and convenient Network of Chips(NoC\cite{NOC1,NOC2}. Due to the highly configurable and adaptive characteristics of AIE and the support of PL, ACAP architecture obtains a finer design granularity than GPU, which can further design hardware level parallelism and pipelining for specific applications. At the same time, compared with FPGA and ASIC, ACAP architecture can achieve a smaller and better design space. Feasible solutions or even optimal solutions for deploying accelerators are obtained faster.

\begin{table}[!t]
\caption{Comparison of Transformer Accelerators on Different Platforms.}
\centering
\begin{tabular}{cccc}
\hline
Platform&GPU&FPGA&Versal ACAP\\
\hline
\makecell{Prefer\\Universality} & \makecell{CUDA\cite{cuda}\\ROCm\cite{rocm}} & \makecell{Auto-ViT-Acc\cite{FPGA_ACC_AUTOVITACC}\\\makecell{Siyuan Lu\\et al.\cite{Siyuan_Lu_et_al}}\\\makecell{Haikuo Shao\\et al.\cite{Haikuo_Shao_et_al}}} & \makecell{SSR\cite{SSR}\\CHARM\cite{CHARM}} \\

\hline

\makecell{Prefer\\Performance} & \makecell{Flash\\Attention\cite{flashattention,flashattention2}} & \makecell{ViA\cite{FPGA_ACC_VIA}\\ME-ViT\cite{FPGA_ACC_MEVIT}} & \textbf{Ours work} \\
\hline
Power(W) & $10^2$ & $10^1$ & $10^1$ \\
\makecell{Throughput\\} & $10^{14}$ Int8& $10^{12}$ Int8 & $10^{14}$ Int8 \\
\hline
\end{tabular}
\end{table}

In Table I, we summarize the representative works of Transformer accelerators on three platforms: GPU, FPGA, and Versal ACAP, and classify them according to the pursuit of generality and performance. High versatility means that the accelerator can adapt to different models of inference with few changes, but this design may not be the best fit for hardware. High performance means that the accelerator specifically customizes the accelerator by considering the upper model and hardware characteristics. This design can often achieve higher performance, but it is inferior in universality. In GPU and FPGA, whether it is inclined to generality or inclined to performance, there are corresponding research and development libraries, because these two kinds of hardware are more classical and mature. In Versal ACAP architecture, SSR\cite{SSR} and CHARM\cite{CHARM} put forward two general frameworks. SSR builds the accelerator by building multiple computing units of the same architecture, and uses spatial sequential hybrid scheduling computing units on the top layer. CHARM proposed an automatic generation framework for Matrix Multiplication (MM) accelerator for Versal ACAP architecture, so as to accelerate the upper model. However, these two works tend to achieve higher generality, and the side effect of this advantage is also obvious, which will sacrifice some performance. Therefore, it is still worth exploring how to achieve better high-throughput and low-power Transformer accelerators on Versal ACAP.

To this end, we propose the CAT framework to fill the gap that favors customized high-performance Transformer accelerators under the Versal ACAP architecture, and jointly advance the ACAP architecture with other works.

\section{CAT Framework Architecture}
\subsection{Overall Framework Architecture}

\begin{figure}[!t]
\centering
\includegraphics[width=3.5in]{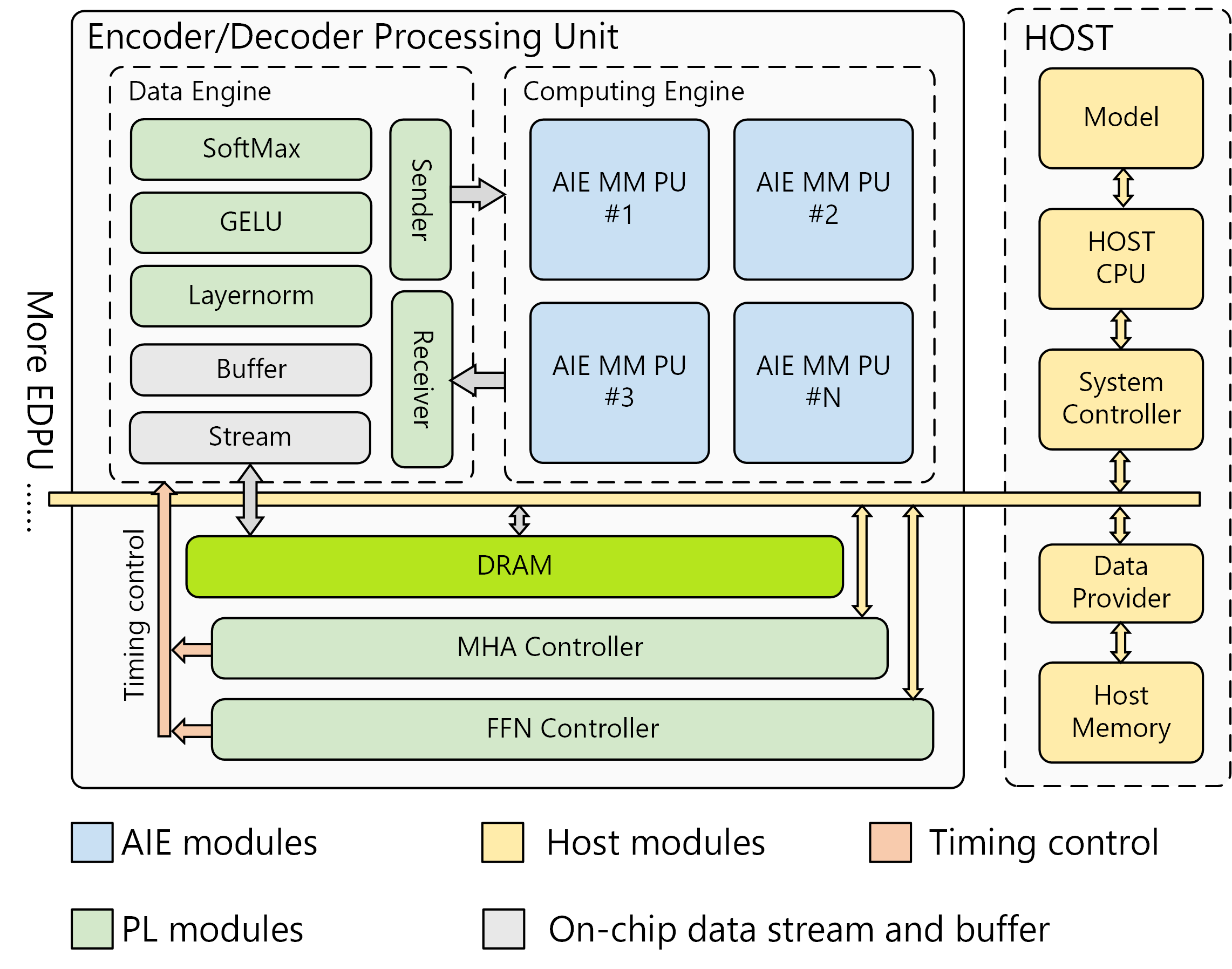}
\caption{The accelerator top-level architecture of the CAT framework}
\label{fig_2}
\end{figure}

\begin{figure*}[!t]
\centering
\includegraphics[width=1.0\textwidth]{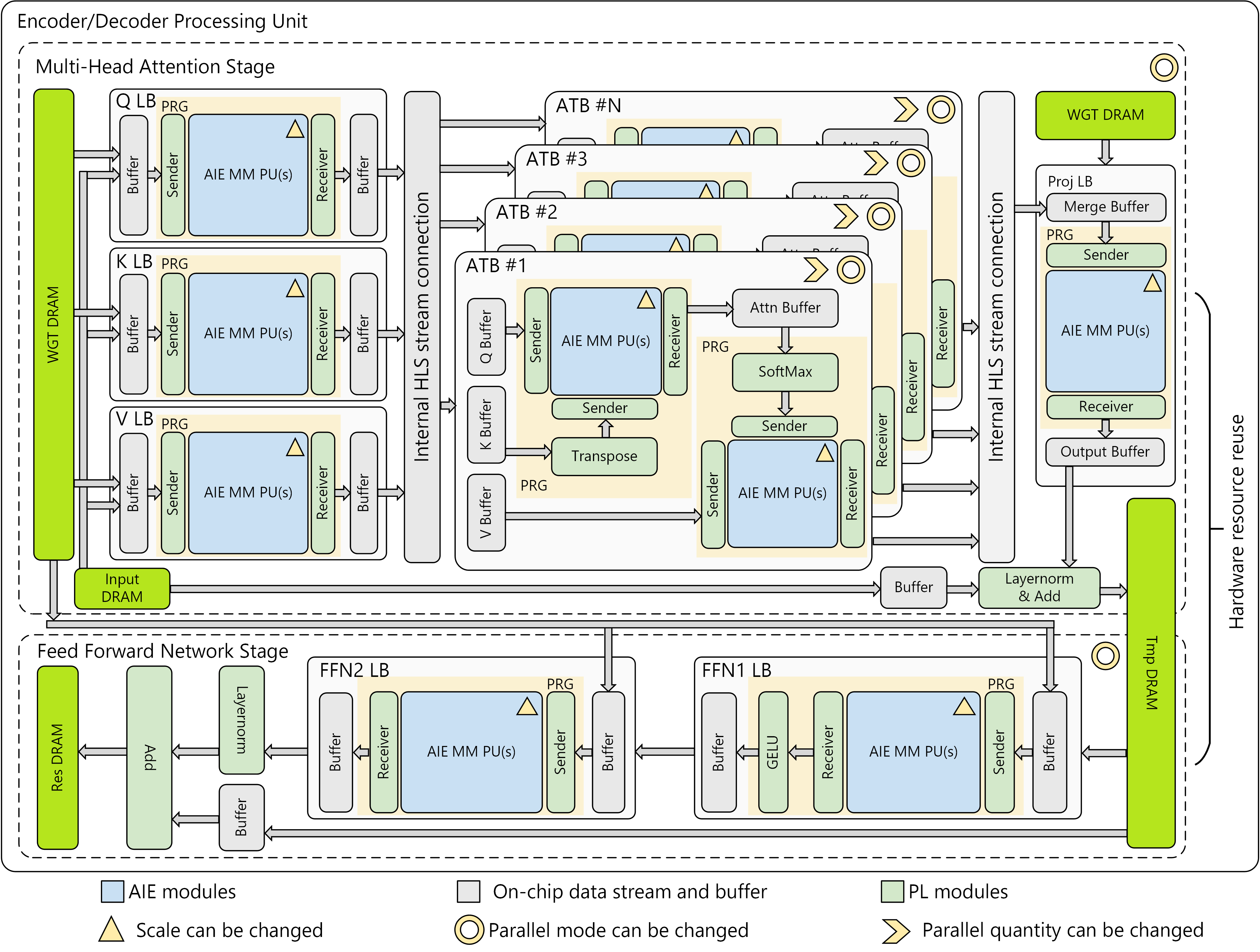}
\caption{EDPU Architecture.}
\label{fig_3}
\end{figure*}

The goal of CAT framework is to use all AIE computing resources as effectively as possible to accelerate the model inference process. We combine the architectural characteristics of Versal ACAP with the characteristics of Transformer, and design the accelerator deployment scheme with Encoder/Decoder as the core. On the one hand, according to the characteristics of AIE, a highly flexible and highly parallel computing engine is designed to fully demonstrate the configurability and super computing power of AIE. On the other hand, in order to meet the needs of the computing engine, the PL side custom-designed data engine.

The accelerator top-level architecture of CAT framework is shown in Figure 2. Encoder/Decoder is implemented in ACAP as an atomic acceleration Unit, and the framework consists of Encoder/Decoder Processing Unit(EDPU) and HOST. One call of the EDPU can complete one operation of the Encoder/Decoder layer. At the same time, the framework supports the deployment of multiple EDPUs to be scheduled in hardware. Different EDPUs can be used to jointly accelerate one upper level task in a pipelined manner, or multiple upper level tasks can be executed in parallel without interfering with each other. The number of EDPUs can be adjusted freely according to hardware resources and acceleration requirements.

The internal structure of EDPUs is highly customizable. We need to combine the framework customization strategy with the Transformer model itself to decide the best organization mode inside EDPUs in a "top-down" way, so the parallelism, organization, data flow connection, etc. of the internal components of EDPUs can be customized. This section only introduces the types and relationships of components inside EDPU.

EDPU consists of a data engine, a computing engine, a dedicated DRAM memory and a controller. The computing engine contains one or more groups of AIE Matrix multiplication processing unit(AIE MM PU), which is responsible for the completion of large-scale and large number of matrix multiplication (MM) operations. The data engine has strong data operation ability and high-speed on-chip cache, which is used to meet the data exchange requirements of the computing engine (Sender, Receiver), and undertake the operation of various nonlinear operators (Softmax, GELU, Layernorm). In addition, due to the precious storage resources on PL chip, we read task data and store results through DRAM, and the whole system uses DRAM as the data exchange center to realize the data exchange of massive parameters and intermediate calculation results.

An execution of EDPU consists of two stages, Multi-Head Attention(MHA) stage and Feed Forward Network(FFN) stage. The parallelization scheme of these two stages makes decisions according to the upper model (e.g., serial, data parallel, pipeline parallel, hybrid mode, etc.). Therefore, we cannot use a fixed scheduling strategy to control the computing engine and data engine. We need to schedule various components by MHA Controller and FFN Controller at PL after the internal structure of EDPU has been determined according to the upper model.

HOST is used to control the top-level resource scheduling of the framework and the execution timing control of EDPUs. It is only responsible for the scheduling work between EDPUs, and cannot interfere with the internal operation of EDPUs. The CPU of the HOST in the hardware side transfers the data required by the Runtime to the DRAM of the hardware accelerator through PCIE. At the same time, the storage space allocation and the start/stop of the hardware accelerator are controlled by the Xilinx Runtime(XRT) and the HOST controller in the software side.

\subsection{EDPU Design}

EDPU is the core component of the accelerator architecture. Figure 3 shows the EDPU architecture. One call of EDPU can complete a Transformer Encoder/Decoder Layer operation. At runtime, it is divided into two stages: Multi-Head Attention(MHA) Stage and Feed Forward Network(FFN) Stage. Based on the available hardware resources and the upper model, the parallel scheme adopted in each stage is determined jointly. The two stages are fixed to serial execution, and the data flow and resources are reorganized for different stages at runtime. The two stages share hardware resources to avoid idle hardware resources. EDPU is mainly composed of Attention Block(ATB) and Liner Block(LB), which are connected by internal data flow, and the path of data flow may contain buffers and nonlinear operators. ATB and LB also have on-chip buffers and nonlinear operators, and contain the AIE Matrix Multiplication Processing Unit(AIE MM PU) with high throughput to complete large-scale MM operations.

In the running process of EDPU, MHA Stage and FFN Stage execute serially and share hardware resources, and the internal of each Stage needs to decide the parallel mode according to the available hardware resources and the upper model to be accelerated to maximize the hardware level. PRG(Parallel Region) is the smallest unit of EDPU scheduling, and each PRG is marked with orange background in Figure 3. ATB and LB may contain one or more PRGS, and the internal fixed parallel pipeline of PRG can ensure that PRG can run at maximum efficiency. No additional AIE MM PU resources will be occupied because there are only PL-end modules in each PRG except for a set of AIE MM pus (s). Although the internal architecture of the minimum scheduling unit PRG is fixed, multiple PRGS can be combined in different parallel ways, which provides a great space for the customization of the framework. This enables us to design the framework under three customization attributes: AIE MM PU scale customization, parallel mode customization, and ATB parallel number customization. In order to effectively optimize the design solution space without losing flexibility, we use triangle, circle and arrow to represent the scope of the three customization attributes in the framework in Figure 3.

\newcommand{\INDSTATE}[1][1]{\STATE\hspace{#1\algorithmicindent}}
\begin{algorithm}[t]
    \caption{EDPU Execution Process.}
    \label{alg:AOA}
    \renewcommand{\algorithmicrequire}{\textbf{Input:}}
    \renewcommand{\algorithmicensure}{\textbf{Output:}}
    \begin{algorithmic}[1]

        \REQUIRE $AIE\_MM\_PUs$, $Parallel\_mode$, $PRGs$, $batch\_size$
        \ENSURE $ResultMemoryBank$    

        \STATE $\textbf{Hardware\_Module MHA\_Stage\{}$
        \STATE $MPM = Parallel\_mode.MHA();$
        \STATE $AIE = AIE\_MM\_PUs.get();$
        \FOR{each $b \in [1,batch\_size]$}
            \STATE //Allocate resources and schedule for PRGs according to parallel mode
            \STATE $\textbf{MHA\_Parallel\_mode\{}$
              \STATE $PRGs.Q\_LB\_prg.run(AIE.Allocate(MPM));$
              \STATE $PRGs.K\_LB\_prg.run(AIE.Allocate(MPM));$
              \STATE $PRGs.V\_LB\_prg.run(AIE.Allocate(MPM));$
              \FOR{each $i \in [1,PRGs.get\_ATB\_prgs\_size()]$}
                \STATE $PRGs.ATB\_prgs[i].run(AIE.Allocate(MPM));$
              \ENDFOR
              \STATE$ PRGs.Proj\_LB\_prg.run (AIE.Allocate(MPM));$
              \STATE$ Layernorm\_Add.run(ResultMemoryBank);$
            \STATE $\textbf{\}}$
          \ENDFOR
       \STATE $\textbf{\}}$

        \STATE $\textbf{Hardware\_Module FFN\_Stage\{}$
        \STATE $FPM = Parallel\_mode.FFN();$
        \STATE $AIE = AIE\_MM\_PUs.get();$
        \FOR{each $b \in [1,batch\_size]$}
            \STATE $\textbf{FFN\_Parallel\_mode\{}$
              \STATE $PRGs.FFN1\_LB\_prg.run(AIE.Allocate(FPM));$
              \STATE $PRGs.FFN2\_LB\_prg.run(AIE.Allocate(FPM));$
              \STATE$ Layernorm\_Add.run(ResultMemoryBank);$
            \STATE $\textbf{\}}$
          \ENDFOR
       \STATE $\textbf{\}}$
       
        \STATE $\textbf{Hardware\_Module EDPU\{}$
           \INDSTATE $\textbf{Serial\_mode\{}$
           \INDSTATE $MHA\_Stage.run();$
            \INDSTATE $FFN\_Stage.run();$
        \INDSTATE $\textbf{\}}$
        \STATE $\textbf{\}}$
        
    \end{algorithmic}
\end{algorithm}

In Algorithm 1, MHA Stage first obtains its parallel organization mode and plans the overall resources, and then calls multiple PRGs in its internal ATB and LB by the specified parallelization method, and allocates AIE computing resources for it by the computing engine at runtime. If we use the pipeline parallel execution mode, all modules within the scope of MHA\_Parallel\_mode will be started at the same time, forming a pipeline as a whole, and each PRGs can also be allocated its own AIE MM PU resources. If we use the serial execution mode, each module will be executed in order. Each PRGs uses AIE MM PU resources in turn, and regardless of the execution mode, MHA Stage should utilize all the computing resources of the computing engine. Finally, the specified batch\_size number of times will be looped over to achieve multi-batch processing. The operation mechanism of FFN Stage is similar to that of MHA Stage, except that it has different resources. Meanwhile, FFN Stage can also adopt a completely different parallel mode from MHA Stage.

MHA Stage and FFN Stage execute serially and share hardware resources. On the one hand, when the EDPU processes the same model, the two stages need the output results of the other side, and there will be a Stage in the blocking state when running, so even if the two stages are parallelly implemented, it cannot guarantee that all AIE cores are running. On the other hand, if the MHA Stage and FFN Stage are also pipelined in parallel, the FFN Stage is in a position with deep pipeline depth, and the start-up time is longer, resulting in a shorter time for the pipeline to run at full speed.

As can be seen from Figure 3, we separate the three QKV linear layers of multiple attention heads from the calculation of attention. This is because after separating the linear layers, the size is usually small due to the head splitting, and it may need to be filled when it is calculated in AIE MM PU. However, the QKV linear layer itself supports extraction and operation together, so we choose to extract and aggregate the small QKV calculations of each attention head into a whole, and use a larger AIE MM PU for operation, which fully ensures the AIE utilization and higher PLIO data reuse rate, which is also of great help to reduce the PLIO load. This linear layer operation method can calculate the amount of data required by multiple head attention at one time, and send the calculation results to the on-chip buffers of multiple ATBs in parallel through the internal data flow, so that the next round of calculation can be directly started without waiting for the ATBS to complete the calculation.

With the same number of AIE cores utilized by EDPU, when the size of MM operation is larger than the core size of AIE allocated by AIE MM PU, the performance of the serial execution (temporal architecture) module will be higher than that of the pipelined (spatial architecture) module, because the serial execution module can use all AIE cores. Pipelined modules need to add more data flow on top of that. When the scale of MM operation is smaller than the core scale of AIE allocated by the AIE MM PU, the disadvantages of the temporal architecture module will appear, and then the pipelined spatial architecture module is needed. Using multiple small AIE MM pus in parallel can achieve lower resource waste, so as to make full use of AIE resources. Although this will lead to an increase in the number of PLIO uses, the small AIE MM PU is not used globally within the entire accelerator, so this effect will be greatly diluted. In addition, in order to improve the efficiency of data exchange from the data engine to the computing engine, we equip each AIE MM PUs with a special Sender and Receiver at the PL side to ensure the parallel sending and receiving of data.

\begin{table}[!t]
\caption{The table compares the operation efficiency of different ways.}
\centering
\begin{tabular}{cccc c}
\hline
ID& \makecell{Independent\\Linear} &\makecell{ATB\\Parallel Mode}&\makecell{ATB\\Parallelism}&\makecell{Speedup\\Ratio}\\
\hline
Lab 1&\ding{53}&N/A&1&1.0x(baseline)\\
Lab 2&\ding{53}&Pipeline Parallel&1&3.8x\\
Lab 3&\ding{52}&N/A&4&5.3x\\
Lab 4&\ding{53}&Pipeline Parallel&4&14.6x\\
Lab 5&\ding{52}&Pipeline Parallel&4&20.1x\\
\hline
\end{tabular}
\end{table}
When AIE cores are fully available, the EDPU's fully parallel unrolled state (Stage internal pipelined parallel + multiple ATB parallel +ATB internal pipelined parallel) yields the highest performance because it makes use of as many cores as possible. However, when the AIE core is not sufficient relative to the task size, other customized states that do not adopt the fully parallelized state can also fully utilize AIE. We believe that this EDPU design scheme and AIE utilization strategy are suitable for ACAP architecture, and give full play to the advantages of ACAP architecture. To prove this, we adopt the model configuration of ViT-Base and design five groups of experiments for three indicators: whether QKV linear layer is integrated with ATB, different architecture paradigms of ATB, and ATB parallelism. To ensure fairness during testing, we all use the same scale AIE MM PU for experiments, and the experimental results are shown in Table II.

As can be seen from Table II, Lab 4 achieves 14.6 times speedup compared to Lab 1 benchmark due to the intervention of pipeline parallelism and multi-ATB parallelism. In Lab 3, the speedup is not obvious because the data generated by the linear layer cannot be consumed in time due to the non-pipelining of serial ATB, resulting in blocking. In Lab 5, it achieves 20.1 times speedup compared with the baseline. The experimental results show that these customized attributes have a huge impact on the actual performance of ACAP architecture, so whether the decision of customized attributes is reasonable is crucial.

\subsection{AIE Evaluating Indicator}

In order to more fully evaluate whether AIE is more fully utilized, we introduce two evaluation indicators: $AIE_{deployment\_rate}$ and $AIE_{effective\_utilization\_rate}$, and the calculation methods are shown in Equation 1, Equation 2. Where $AIE_{Total\_number}$ is the total number of AIE resources, $AIE_{Deployment\_number}$ is the number of AIE deployments, and $AIE_{Running\_number}$ is the number of effective AIE runs.

\begin{equation}
\label{eq1}
AIE_{deployment\_rate}=\frac{AIE_{Deployment\_number}}{AIE_{Total\_number}}
\end{equation}

\begin{equation}
\label{eq2}
AIE_{effective\_utilization\_rate}=\frac{AIE_{Running\_number}}{AIE_{Deployment\_number}}
\end{equation}

$AIE_{deployment\_rate}$ is the ratio of deployed AIE to the total number of AIE resources, which can represent the actual number of deployed AIE in the board. The deployed AIE has the ability to take on tasks and can wait to be called. The $AIE_{effective\_utilization\_rate}$ of AIE is the ratio of running AIE to deployed AIE, which can represent the actual running number of AIE at a certain time or stage. After distributing tasks by the data engine, the deployed AIE will be transformed into running state when it effectively assumes the task amount.

We distinguish the number of AIE deployments from the number of effective AIE runs. This is because in complex tasks (such as Transformers), the hardware may run in many stages or utilize different hardware resources at different points in time, often resulting in a part of the deployed AIE cores not being called, or in a blocked state. This kind of AIE core is obviously not helpful to improve the performance, so the $AIE_{effective\_utilization\_rate}$ metric, which can judge the number of effectively running AIE, is very important. In addition, the $AIE_{deployment\_rate}$ is also an important indicator, which determines the upper limit of the performance of the system. If the $AIE_{deployment\_rate}$ is low, even the excellent $AIE_{effective\_utilization\_rate}$ indicator cannot obtain higher performance.

\section{CAT Framework Customization and Optimization Strategy}
\subsection{Transformer Load Analysis}

The configuration information of Transformer model usually includes the number of heads ($Head$), the dimension of embedding vector ($Embed\_dim$) and the dimension of FFN hidden layer ($Dff$), and the length of input sequence ($L$). The parameters that determine the calculation amount of multi-head attention module are $Head$, $Embed\_dim$ and $L$. When calculating each Head of attention separately, it first needs to go through three linear layers of QKV of size$ [Embed\_dim, Embed\_dim/Head]$. Then, matrix multiplication (MM) was performed between Q matrix of size $[L, Embed\_dim/Head]$ and the transpose of K matrix of size $[Embed\_dim/Head, L]$, and the weights of size $[L, L]$ were obtained by softmax function. Then, this weight is combined with the V matrix of size $[L, Embed\_dim/Head]$ to do MM operation. Finally, the results of each head are aggregated, and then the output of the multi-head attention module is obtained by the Projection linear layer of size $[Embed\_dim, Embed\_dim]$. The parameters that determine the calculation amount of the feedforward network are $Embed\_dim$ and $Dff$. Firstly, the output of the multi-head attention module needs to be passed through a linear layer of size $[Embed\_dim, Dff]$. The output of the linear layer is processed by the GELU function, and then passed through the linear layer of size $[Dff, Embed\_dim]$ to obtain the final output.

It can be drawn that computing a MHA and a FFN requires $5×Head+3$ matrix multiplications, $Head$ softmax and $Head$ matrix transpose, among which MM operations have the characteristics of large number and small scale, and only three MM operations are large-scale, even after the QKV linear layer is extracted, there are also a large number of small MM operations. Therefore, it is often inefficient to only accelerate the MM operator and call it multiple times, because these small MM operations are difficult to use all the hardware resources, resulting in a waste of resources. Therefore, the highly customizable characteristics of Versal ACAP architecture and operator fusion technology are fully utilized to map these operations into hardware efficiently, which is helpful to improve efficiency and increase hardware resource utilization.

\subsection{AIE MM PU Customized Design}

AIE MM PU is the most important part of the computing engine and plays a major role in the operation process of the whole system. At the same time, AIE MM PU of different scales are required in MHA Stage and FFN Stage, so how to design efficient AIE MM PU for the hardware resources of the board is crucial. This section describes the customization attributes of AIE MM PU scale customization in EDPU. Table III shows the symbol information used in the paper and its meaning.

\begin{table}[!t]
\caption{The table compares the operation efficiency of different ways.}
\centering
\begin{tabular}{ccc}
\hline
Categories&Symbol&Meaning\\
\hline
\multirow{4}{*}{\makecell{Transformer model\\configuration\\information}}&$Head$&Head of attention\\
~ & $Embed\_dim$ & Embedding dimension\\
~ & $Dff$ & Feed Forward dimension\\
~ & $L$ & Input sequence length\\
\hline
\multirow{8}{*}{\makecell{Intrinsic\\hardware\\parameters}}&$T_{PU}$&AIE MM PUIteration time\\
~ & $T_{Window}$ & \makecell{PLIO takes time to transmit\\ AIE Window once}\\
~ & $T_{Calc}$ & AIE single-core iteration time\\
~ & $MMSZ_{AIE}$ & AIE single core load size\\
~ & $bit_{data}$ & The bit width of the data\\
~ & $M_{Window}$ & AIE Window size (bytes)\\
~ & $Total_{AIE}$ & Number of all AIE resources\\
~ & $Total_{Buffer}$ & All on-chip storage space\\
\hline
\multirow{3}{*}{\makecell{Configurable\\parameters}}&$PM_{MHA}$&MHA Stage Parallel mode\\
~ & $PM_{FFN}$ & FFN Stage Parallel mode\\
~ & $P_{ATB}$ & ATB Degree of parallelism\\
\hline
\end{tabular}
\end{table}

In the design of AIE MM PU, it is necessary to consider the load of AIE single core and the overall load of AIE MM PU. For the load of AIE single core, we suggest to choose the matrix multiplication size of square (${MMSZ_{AIE}}^3$). Because this facilitates the expansion of core size when using block parallel matrix multiplication, AIE vector processor characteristics (instruction length is a power of 2), AIE Window capacity (storage space for AIE core input and output), determine the AIE single-core load. We use Equation 3 to constrain the payload size of AIE single core.

\begin{equation}
\label{eq3}
\left\{ \begin{matrix}
{{{MMSZ}_{AIE}}^{2} \times {bit}_{data} \leq M_{Window}/4} \\
{{MMSZ}_{AIE} \in \left\lbrack 1,2,4,\cdots 2^{n} \right\rbrack\left( n \in N^{*} \right)}
\end{matrix} \right.
\end{equation}

We set the number of bytes consumed by the AIE single-core workload to be less than or equal to 1/4 of the AIE Window capacity, because an AIE requires two Windows for input and output, and double buffering, which fully utilizes the entire AIE Window capacity. Meanwhile, the instruction length of AIE vector processor is a power of 2, so the $MMSZ_{AIE}$ should be set to a power of 2 for convenient calculation.

Aiming at the overall load of AIE MM PU, more cores should be organized, which will bring greater data communication demand, so the number of AIE single cores in AIE MM PU organization should be determined by considering the efficiency of PLIO data communication. $T_{Window}$ is the time taken by PLIO to transmit one AIE Window. 
When PLIO is multiplexed using Packet Switch mode, the transmission time of one PLIO is $PLIO_{AIE}$×$T_{Window}$ when it is responsible for $PLIO_{AIE}$ single core. 
Therefore, We need to make this transmission time less than $T_{Calc}$ so that all
cores inside AIE MM PU do not lose efficiency, that is, $T_{PU} \approx T_{Calc}$, and $PLIO_{AIE}$ becomes the maximum size of 2D core group that AIE MM PU can expand. $PLIO_{AIE}$ can be constrained by Equation 4.

\begin{equation}
\label{eq4}
{PLIO}_{AIE} \leq \left\lfloor \begin{matrix}
{T_{Calc} \div T_{Window}}
\end{matrix} \right\rfloor
\end{equation}

Thus, the maximum number of cores allowed by an AIE MM PU containing AIE is ${{PLIO}_{AIE}}^2$.

\begin{figure*}[!t]
\centering
\includegraphics[width=1\textwidth]{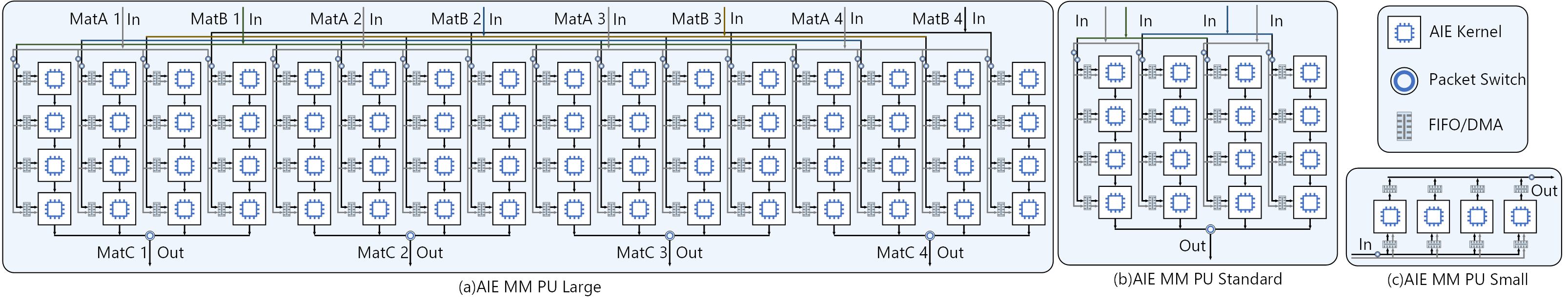}
\caption{AIE MM PU design with different specifications.}
\label{fig_3}
\end{figure*}

Using the above constraints, we design three kinds of AIE MM PU for the VCK5000 board with 400 AIE cores, as shown in Figure 4, which are Large(a) and Standard(b), Small(c), respectively.  $PLIO_{AIE}$=4.

The Large size PU contains 64 AIE cores, 8 input PLIO channels and 4 output PLIO channels, and completes the matrix multiplication task of 4$MMSZ_{AIE} \times 4MMSZ_{AIE} \times 4MMSZ_{AIE}$ at one time. The Standard PU contains 16 AIE cores, including 4 input PLIO and 1 output PLIO, which can complete the matrix multiplication task of 2$MMSZ_{AIE} \times 4MMSZ_{AIE} \times 2MMSZ_{AIE}$ at one time. Small size PU contains four AIE cores, two input PLIO and one output PLIO, and completes the matrix multiplication task of $MMSZ_{AIE} \times MMSZ_{AIE} \times 4MMSZ_{AIE}$ at one time.

When designing the accelerator, the AIE MM PU module in Figure 3 can select the appropriate AIE MM PU specification according to the Transformer model specification to be accelerated, so as to achieve higher performance and make more rational use of hardware resources.

\subsection{Parallel Mode Customized Design}
Parallel mode is the most important customization attribute of EDPU, which determines the way of resource allocation and organization of the internal modules of the two stages. Whether the decision of parallel mode is reasonable directly affects the performance of the hardware. Therefore, how to combine the characteristics of the upper Transformer and the hardware resources of the board to set a reasonable parallel mode is crucial. This subsection describes two aspects: how to customize the parallel mode and the parallelism decision of ATB.

We take PRG(Parallel Region) as the basic unit of discussion. Because its internal fixed adopts pipeline parallelism, it can hide the internal details of PRG and discuss the parallel mode of PRG's upper layer (between PRGS).

We present two optimal parallel modes: (1) fully pipelined parallelization mode. (2) serial-parallel hybrid mode with serial transmission but ATB parallel transmission. In (1), all PRGs are launched in parallel, each PRG exclusively occupies a part of the resources in the computing engine (AIE MM PUs), and the sum of the resources of each PRG is equal to all the resources of the computing engine. The whole Stage forms a large pipeline and completes the calculation at one time. In (2), LBs are allowed to transmit multiple ATBs in parallel, and the two PRGS of ATB are serial, and the computing resources are equally distributed among multiple ATBs. Firstly, all resources were used to execute the three QKV LB in series, and then multiple ATBs were launched in parallel to process the multi-head attention in parallel. Finally, the calculation was completed by executing the Proj LB.
In addition, pure serial execution mode is only tried if the scale of all MM operations for the Transformer model (including the small MM in the MHA stage) is larger than the scale of the computation engine as a whole at once, but this is extremely rare and therefore not discussed.

We jointly decide the parallel mode inside Stage from three aspects: the internal operator characteristics of Transformer model (Transformer model configuration information in Table III), the maximum matrix multiplication scale that AIE MM PU can undertake, and whether the PL on-chip resources are enough. We use the formula 5 to constrain the decision of the parallel mode of the MHA Stage($PM_{MHA}$).

\begin{equation}
\label{eq5}
\left\{ \begin{matrix}
{Factor1 = \frac{L \times {Embed\_ dim}^{2}}{\left. {\left\lfloor {{Total}_{AIE}/{{PLIO}_{AIE}}^{2}} \right\rfloor \times \left( PLIO \right.}_{AIE} \times {MMSZ}_{AIE} \right)^{3}}} \\
{Factor2 = {\sum\limits_{1}^{n}{LB.buf\lbrack n\rbrack.size}} + {\sum\limits_{1}^{m}{ATB.buf\lbrack m\rbrack.size}}} \\
{Factor1 \geq {PRG}_{MAX\_ Pipeline\_ Depth}} \\
{Factor2 > {Total}_{Buffer}}
\end{matrix} \right.
\end{equation}

In formula 5, two factors are used to jointly determine the parallel mode of MHA Stage, where Factor1 represents the ratio of the MM scale required by the Transformer model in LB to the maximum MM scale that the computing engine can undertake at one time. Factor2 represents the PL on-chip storage space consumed by MHA Stage under the condition of fully pipelined unrolling. When the value of Factor1 is greater than ${PRG}_{MAX\_Pipeline\_Depth}$, or the value of Factor2 is greater than the sum of PL on-chip memory space, Parallel Mode (2) should be used. In other cases, Parallel Mode (1) should be used to make the best use of hardware resources.
begin{equation}

\begin{equation}
\label{eq6}
\left\{ \begin{matrix}
{Factor1 = \frac{L \times Embed\_ dim \times Dff}{\left. {\left\lfloor {{Total}_{AIE}/{{PLIO}_{AIE}}^{2}} \right\rfloor \times \left( PLIO \right.}_{AIE} \times {MMSZ}_{AIE} \right)^{3}}} \\
{Factor2 = FFN1_{LB}.buf\lbrack n\rbrack.size + FFN2_{LB}.buf\lbrack n\rbrack.size} \\
{Factor1 \geq {PRG}_{MAX\_ Pipeline\_ Depth}} \\
{Factor2 > {Total}_{Buffer}}
\end{matrix} \right.
\end{equation}

For FFN Stage, we use formula 6 to constrain the decision of the parallel mode of FFN Stage($PM_{FFN}$), which is the same as MHA Stage, and only modified at the target MM scale of Factor1. Factor2 is changed to the PL on-chip storage space consumed by the two LB of FFN Stage under the condition of fully pipelined expansion, and the decision conditions for Factor1 and Factor2 are also the same.

\subsection{ATB Parallelism Customized Design}
Since the ATB only exists in the MHA Stage, after the $PM_{MHA}$ is determined, the throughput of LB is also determined, so we can make a decision on the customization property of ATB parallelism ($P_{ATB}$). We have two ways to decide $P_{ATB}$.

On the one hand, if the number of results output by QKV LB in one execution is integer proportional to the data consumed by ATB in one execution, this ratio can be directly adopted to determine $P_{ATB}$, and we can use the formula 7 to constrain $P_{ATB}$.

\begin{equation}
\label{eq8}
{P_{ATB} = QKV\_ LB}_{Output\_ Heads}/{ATB}_{Input\_ Heads}
\end{equation}

On the other hand, if the amount of data exchanged between LB and ATB is not integer proportional, the ratio can be calculated from the perspective of throughput, thus determining $P_{ATB}$, and we can use the formula 8 to constrain $P_{ATB}$.

\begin{equation}
\label{eq8}
{P_{ATB} = Throughput}_{QKV\_ LB}/{Throughput}_{ATB}
\end{equation}

\subsection{Versal ACAP Optimization Strategy}

In Versal ACAP performance optimization strategy, we choose to adopt the idea proposed by EA4RCA framework\cite{EA4RCA}, which is a preliminary work done by our team on Versal ACAP, aiming to maximize the performance of AIE in Communication avoiding(CA)\cite{CA} applications. Based on the above work, this paper adopts the following three optimization strategies to optimize the hardware design and design process, so as to further improve the performance of the CAT framework:

\begin{itemize}
\item{\textbf{Data transmission strategy for decoupling computation and communication}: The MM load inside the Transformer is borne by the AIE. MM belongs to the category of CA, so we take advantage of its easy decoupling of computation and communication to reduce the number of AIE computation interruptions, while making full use of the DMA engine inside AIE, which is able to significantly improve the performance.}
\item{\textbf{The hardware design method of computing engine and data engine guided by "top-down" }: using the "top-down" design method starting from the characteristics of the application, so that developers can quickly obtain relatively optimal solutions, faster to get feasible solutions. At the same time, due to the introduction of the concept of computation engine and data engine, different modules can be deployed and implemented on the hardware architecture they are good at, and the optimization methods proposed by EA4RCA framework can be easily applied when designing EDPU.}
\item{\textbf{AIE Graph Code Generator Optimizes and standardizes AIE development process}: When writing customized AIE MM PU of different scales in EDPU, we generate compilable AIE engineering code of AIE MM PU in the calculation engine with one click by importing configuration files, which greatly improves the AIE development efficiency and reduces the development difficulty.}
\end{itemize}

\section{Evaluation}

We evaluate the CAT framework on two typical Transformer models: BERT and ViT. Our evaluation shows that the CAT framework can effectively customize hardware accelerators according to the characteristics of the hardware and the characteristics of the model itself, with faster inference speed while maintaining high hardware utilization.

\subsection{Experimental Setup}

\textbf{Hardware Setup:} We implement the accelerator on a hybrid system of CPU-ACAP. The system was deployed on a machine equipped with an Intel Xeon Gold 6148 2.40GHz processor, 64GB main memory, and 1 TB SSD hard disk. The Versal ACAP Card is the Xilinx VCK5000 Versal Development Card, which is built on the Xilinx 7 nm Versal ACAP architecture and provides 145TOPS Int8 peak performance. On VCK5000, the on-chip SRAM capacity is 23.9MB, the bandwidth is 23.5TB /s, the off-chip memory capacity is 16GB, and the total off-chip bandwidth is 102.4GB /s. Data transfer between the Host CPU and the VCK5000 is realized through the PCI Express interface.

\textbf{Software Setup:} We used Xilinx Vitis 2022.2 toolchain to build the software environment and AMD Power Design Manager (PDM) 2023.2.2 to evaluate the device power consumption. In terms of hardware construction, we use aiecompiler tool to build AIE hardware, use v++ tool to build PL kernel and perform hardware connection construction, so as to complete the generation of Xilinx binary container file (xclbin). In terms of hardware simulation, aiesimulator is used to complete the performance and correctness analysis of AIE, and Vitis analyzer is used to view the simulation results. For the actual operation, we control the operation of the board using the Xilinx Runtime(xrt), which is the software interface to the Xilinx programmable logic device.

\textbf{Benchmarks:} We select two typical Transformer models to verify the effectiveness of the CAT framework: (1) BERT-Base: It is a model in NLP domain that consists of a multi-layer bidirectional Transformer Encoder architecture, whose sequence length (L) is fixed to 256 in the experiments of this paper. (2) ViT-Base: It is a model in the CV domain that is mainly used to extract image features. For the above models, the already quantified Int8 model is used, which is due to the fact that replacing FP32 with Int8 in Transformer can greatly reduce the computational complexity with limited accuracy loss\cite{8BIT}. Table IV shows the model configuration adopted in our experiments.

In addition, in order to show that the CAT framework can effectively map Transformer models to hardware under different hardware resources, we introduce a set of BERT-Base(Limited AIE) experiments for BERT models, which limits the available number of AIE. It is used to simulate the situation that AIE device resources are different in different Versal ACAP boards.

\begin{table}[!t]
\caption{Configuration information for the three experiments.}
\centering
\resizebox{1.0\columnwidth}{!}{
\begin{tabular}{cccc cccc c}
\hline
Model&Heads&\makecell{Embed\\dim}&Dff&L&Layers&\makecell{Data\\Type}&\makecell{Allowable\\Number\\of AIEs}\\
\hline
BERT-Base&12&768&3072&256&12&Int8&400\\
ViT-Base&12&768&3072&197&12&Int8&400\\
\makecell{BERT-Base\\(Limited AIE)}&12&768&3072&256&12&Int8&64\\
\hline
\end{tabular}
}
\end{table}

\subsection{Design Case Based on CAT}

In this section, we customize the design of the accelerator for the BERT-Base model based on the CAT framework. First, we need to configure the information analysis load for the BERT-Base model in Table IV. Here, we fixed the Independent Linear strategy, one iteration of EDPU, That is, one MHA Stage and one FFN Stage require 4 times of 256×768×768 MM, 12 times of 256×64×256 MM, 12 times of 256×256×64 MM, 2 times of 256×768×3072 MM, 12 times of Softmax and 12 times of matrix transpose.

After that, we map the above Transformer MM load to the AIE MM PU based on the designed AIE MM PU with Large, Standard and Small specifications designed and completed for the VCK5000 board. We assign an AIE MM PU Large to each of the four LB PRGS in the MHA Stage, which is used to undertake four 256×768×768 MM. Two AIE MM PU Small are allocated for the prestage PRG in the ATB to undertake 12 256×64×256 MM. Two AIE MM PU standards are allocated for the post-stage PRG in ATB, which are used to undertake 12 times 256×256×64 MM; Two AIE MM PU Large are allocated to each PRG of the two LB in the FFN Stage to bear 256×768×3072 MM twice.

Thus, both LB and ATB throughput and data exchange ratio in MHA Stage can be calculated, and we can determine the ATB parallelism ($P_{ATB}$). The linear layer of QKV can output 256×256 matrices at a time, while the data required by a single head is 256×64 matrices after attention splitting, so QKV can output the amount of data required by 4 ATBs at a time, so we set $P_{ATB}=4$.

From this, we can determine the parallel mode customization properties of the CAT framework, by the model configuration L=256, Embed\_dim=768, For AIE MM PU properties ${PLIO}_{AIE}$=4, ${Total}_{AIE}$=400, ${MMSZ}_{AIE}$=64 Factor1=1.5 can be calculated from formula 5, and the EDPU architecture shows ${PRG}_{MAX\_Pipeline\_Depth}=4$, $Factor1<{PRG}_{MAX\_Pipeline\_Depth}$.

In addition, the internal on-chip SRAM capacity of the VCK5000 is 23.9MB. When the MHA Stage is fully pipelined, the PL on-chip storage space consumed by the MHA stage is composed of QKV LB output cache, ATB input-output cache, ATB attention cache, Proj LB input-output cache and weight cache. The QKV LB output cache takes up 256×256×3=192KB, the ATB I/O cache takes up 256×64×4×4+=256KB, the ATB attention cache takes up 128×256×4=128KB, and the ATKV LB output cache takes up 256×256× 4= 192KB. The Proj LB input-output cache footprint is 256×768+256×256=256KB, and the weight cache footprint is 768×768×4+768×3072×2=6.75MB. This results in a Total on-chip cache footprint of Factor2=7.5625MB, $Factor2<{Total}_{Buffer}$. Factor1, Factor2 and formula 5 show that the BERT-Base model is suitable for the fully pipelining parallelization mode of the CAT framework.

Finally, considering the characteristics of the hardware itself and the characteristics of the model, we determine three customization attributes of the CAT framework, so that the accelerator can be highly customized for the upper and lower layers, so as to obtain better performance.

\subsection{Hardware Resource Utilization Analysis}

We comprehensively evaluate the hardware resource utilization of hardware accelerators designed based on the CAT framework in terms of AIE utilization, PL on-chip storage resources, and lookup tables. Similarly, for better comparison, we list in Table V the resource occupancy of the three accelerators BERT-Base, ViT-Base and BERT-Base(Limited AIE), and the two utilization metrics of AIE. At the same time, we also analyze three aspects of MHA Stage, FFN Stage and System Overall.

As shown in Table V, taking the BERT-Base accelerator developed based on the CAT framework as an example, 352 AIE cores were deployed in the system as a whole, and the AIE deployment rate reached 88\%. The AIE effective utilization rate in MHA Stage can reach 100\%, that is, all deployed cores are utilized, among which 4 groups of AIE MM PU Large, a total of 256 cores are allocated to 4 LB. The rest of the AIE MM PU is allocated to the ATB, and this part of the AIE MM PU occupies 96 AIEs and is exclusive to the ATB. In the FFN Stage, due to the two-stage hardware resource sharing, the AIE MM PU of the four LB in the MHA Stage is used to complete the calculation, so the AIE effective utilization rate of the FFN Stage is 73\%. The system as a whole also achieves an average AIE effective utilization rate of 87\% at runtime, which achieves high AIE utilization rate. In addition, due to the two-stage hardware resource sharing, the resource consumption of the whole system (EDPU) is not the sum of the data of the two stages, but less than the sum of the individual resource occupancy of the two stages, which is also confirmed by the data in Table V.

\begin{table}[!t]
\caption{Hardware resource utilization of three accelerators designed based on the CAT framework.}
\centering
\resizebox{1.0\columnwidth}{!}{
\begin{tabular}{cccc cccc}

\hline

Model&\makecell{Hardware\\Module}&LUT&FF&BRAM&URAM&\makecell{AIE\\dep.\\rate}&\makecell{AIE\\eff.\\util.\\rate}\\

\hline

\multirow{8}{*}{\makecell{BERT\\Base}} &\makecell{MHA\\Stage}&162.9K&213.6K&588&220&\makecell{88\%\\(352 \\AIEs)}	&\makecell{100\%\\(352\\AIEs)}  \\ 

~&\makecell{FFN\\Stage}&71.7K&85K&482&276&\makecell{88\%\\(352 \\AIEs)}	&\makecell{73\%\\(256\\AIEs)}  \\ 

~& Overall &232.3K&290.5K&940&360&\makecell{88\%\\(352 \\AIEs)}	&\makecell{87\%\\(Avg)}  \\ 

\hline

\multirow{8}{*}{\makecell{ViT\\Base}} &\makecell{MHA\\Stage}&201.6K&221.3K&598&148&\makecell{88\%\\(352 \\AIEs)}	&\makecell{100\%\\(352\\AIEs)}  \\ 

~&\makecell{FFN\\Stage}&71.7K&85K&482&276&\makecell{88\%\\(352 \\AIEs)}	&\makecell{73\%\\(256\\AIEs)}  \\ 

~& Overall &261.4K&262.4K&706&412&\makecell{88\%\\(352 \\AIEs)}	&\makecell{87\%\\(Avg)}  \\ 

\hline

\multirow{8}{*}{\makecell{BERT\\Base\\(Limited\\AIE)}} &\makecell{MHA\\Stage}&46.6K&71.3K&320&0&\makecell{100\%\\(64 \\AIEs)}	&\makecell{100\%\\(64\\AIEs)}  \\ 

~&\makecell{FFN\\Stage}&46.5K&71.2K&320&0&\makecell{100\%\\(64 \\AIEs)}	&\makecell{100\%\\(64\\AIEs)}  \\ 

~& Overall &48.4K&73.1K&320&0&\makecell{100\%\\(64 \\AIEs)}	&\makecell{100\%\\(Avg)}  \\ 

\hline

\end{tabular}
}
\end{table}

Similarly, in the experiment of ViT-Base, the above law is also followed, but because the sequence length L is smaller than that of BERT-Base, the storage resources and PL end resources such as lookup table occupied by its MHA Stage and FFN Stage are relatively low, but the AIE utilization remains the same.

In addition, in the BERT-Base(Limited AIE) experiment, because we limited its AIE scale, most of its parallel mode customization process adopts serial design, avoiding ATB's proprietary AIE core, and each PRGs has the right to have all the computing resources at runtime. Therefore, Its AIE deployment rate and AIE effective utilization rate can both reach 100\%, and the PL on-chip resource consumption caused by serialization is less.

\subsection{Performance and Energy Efficiency Analysis}

In this set of experiments, we analyze the performance and Power consumption of three accelerators developed based on the CAT framework. In order to comprehensively evaluate the actual work of the hardware, we use Latency and TOPS as speed indicators, Power(W) and GOPS/W as power and energy efficiency indicators. It can represent the real performance of the system when the hardware is actually running. These tests are run at the frequency of 1.33GHZ for AIE and 300MHZ for PL.

\begin{table}[!t]
\caption{Peak performance and energy efficiency performance of three accelerators designed based on the CAT framework.}
\centering
\resizebox{1.0\columnwidth}{!}{
\begin{tabular}{cccc ccc}

\hline

Model&\makecell{Hardware\\Module}&\makecell{Latency\\(MS)}&TOPS&GOPS/AIE&\makecell{Power\\(W)}&GOPS/W\\

\hline

\multirow{5}{*}{\makecell{BERT\\Base}} & \makecell{MHA\\Stage} &0.037 &40.237 &\makecell{114.309\\(352 AIEs)} &N/A & N/A  \\ 

~ &\makecell{FFN\\Stage} &0.081 &29.846 &\makecell{116.589\\(256 AIEs)} &N/A & N/A  \\ 
~ &\makecell{System\\(EDPU)} &\textbf{0.118} &\textbf{35.194} &\textbf{\makecell{99.983\\(352 AIEs)}} &\textbf{67.555} & \textbf{520.968}  \\ 

\hline

\multirow{5}{*}{\makecell{ViT\\Base}} & \makecell{MHA\\Stage} &0.049 &30.450 &\makecell{86.505\\(352 AIEs)} &N/A & N/A  \\ 

~ &\makecell{FFN\\Stage} &0.081 &29.846 &\makecell{116.589\\(256 AIEs)} &N/A & N/A  \\ 
~ &\makecell{System\\(EDPU)} &\textbf{0.129} &\textbf{30.279} &\textbf{\makecell{86.020\\(352 AIEs)}} &\textbf{61.464} & \textbf{492.629}  \\ 

\hline

\multirow{5}{*}{\makecell{BERT\\Base\\(Limited\\AIE)}} & \makecell{MHA\\Stage} &0.147 &9.607 &\makecell{150.109\\(64 AIEs)} &N/A & N/A  \\ 

~ &\makecell{FFN\\Stage} &0.252 &9.595 &\makecell{149.922\\(64 AIEs)} &N/A & N/A  \\ 
~ &\makecell{System\\(EDPU)} &\textbf{0.398} &\textbf{9.598} &\textbf{\makecell{149.968\\(64 AIEs)}} &\textbf{16.168} & \textbf{593.642}  \\ 

\hline

\end{tabular}
}
\end{table}

In order to carry out a more comprehensive analysis, in addition to the analysis of the performance and power consumption of the system as a whole, we also analyze the performance of MHA Stage and FFN Stage within each accelerator to show the performance performance more comprehensively. Since we have the full accelerator deployed in hardware, we only provide energy efficiency metrics for the system as a whole (Power and GOPS/W), and the individual performance of the two stages is derived from the runtime reports. In addition, the peak throughput that the system can actually achieve (batch\_size as large as possible) is selected as a reference in the test, and the average is obtained by running multiple times. For different batch\_size, we will describe it in more detail later.

Table VI shows the peak performance and energy efficiency performance of the three accelerators designed based on the CAT framework. In the BERT-Base accelerator, the overall system throughput reaches 35.194TOPS, the overall delay is 0.118 ms, and the energy efficiency is 520.968 GOPS/W. Since the AIE effective utilization rate of MHA Stage is 100\%, MHA Stage can utilize all 352 AIE cores, and its throughput is the highest, reaching 40.237TOPS. At the same time, due to the internal parallelization of full pipeline expansion, the delay of one iteration is very short (0.037ms). For the FFN Stage, because of its large amount of calculation and the AIE effective utilization rate is reduced to 73\%, the throughput is reduced. However, the single-core throughput of AIE of MHA Stage and FFN Stage can reach the same and high level, and the different throughput is only caused by the different number of cores. Therefore, it can be proved that our accelerator is very effective in AIE utilization. In addition, because the two stages share resources and execute serially, the overall performance of the system should be between the two stages, which is also proved by the data in Table VI.

Similarly, for the ViT-Base accelerator, its performance has roughly the same characteristics as the BERT-Base accelerator, except that its MHA Stage throughput is relatively low. At this time, because its sequence length ($L$) is 197, we need to fill it because ${MMSZ}_{AIE}=64$. As a result, a part of the throughput is occupied by the padded data, so its MHA Stage performs worse than the BERT-Base accelerator. Nevertheless, our performance is still ahead of the game.

For the BERT-Base(Limited AIE) experiment, due to the serial operation mode and the small size of the AIE, it can easily release all the performance of AIE and achieve a very high single-core AIE throughput (150 GOPS/AIE). The overall throughput of the system (9.598TOPS) is close to the throughput of 64 AIE cores running only MM (10TOPS), and the efficiency reaches 96\%, which also makes the energy efficiency of AIE more excellent. Therefore, it can be proved that our framework can reasonably plan the parallel mode under different hardware resources to maximize the AIE performance.

\begin{table*}[!t]
\caption{Performance and Energy Efficiency Comparison between CAT Framework and Current SOTA Design.}
\centering
\begin{tabular}{cccc cccc c}

\hline

Model&Platform&Design&Frequency&Precision&\makecell{Throughtput\\(TOPS)}&\makecell{Energy Efficiency\\(GOPS/W)}&\makecell{Speed Up\\Ratio}&\makecell{Energy Efficiency\\Up Ratio}\\

\hline

\multirow{8}{*}{Peak} &NVIDIA A10G&TensorRT\cite{GPU_ACC_A10G}&1.71GHZ&FP32&14.630&66.79&47.35×&8.43×\\
~&Alveo U50&ViA\cite{FPGA_ACC_VIA}&300MHZ&FP16&0.309&7.92&1.00×&1.00×\\
~&ZCU102&Auto-ViT-Acc\cite{FPGA_ACC_AUTOVITACC}&150MHZ&FIX8&0.711&84.10&2.30×&9.36×\\
~&VCK190&\makecell{SSR\cite{SSR}\\(FPGA’24)}&\makecell{AIE:1GHZ\\PL: 230MHZ}&INT8&26.700&453.32&86.41×&57.24×\\
~&Zynq Z-7100&NPE\cite{NPE}&200MHZ&16-bit&0.208&10.40&0.67×&1.31×\\
~&VCK5000&CAT&\makecell{AIE:1.25GHZ\\PL:300MHZ}&INT8&\textbf{35.194}&\textbf{520.97}&\textbf{113.90×}&\textbf{64.78×}\\

\hline

\multirow{6}{*}{ViT} &Alveo U50&ViA\cite{FPGA_ACC_VIA}&300MHZ&FP16&0.309&7.92&1.00×&1.00×\\
~&ZCU102&Auto-ViT-Acc\cite{FPGA_ACC_AUTOVITACC}&150MHZ&FIX8&0.711&84.10&2.30×&10.62×\\
~&VCK190&\makecell{SSR\cite{SSR}\\(FPGA’24)}&\makecell{AIE:1GHZ\\PL 230MHZ}&INT8&22.03&360.04&71.29×&45.46×\\
~&VCK5000&CAT&\makecell{AIE:1.25GHZ\\PL 300MHZ}&INT8&\textbf{30.279}&\textbf{492.629}&\textbf{97.99×}&\textbf{62.20×}\\

\hline

\multirow{3}{*}{BERT} &Zynq Z-7100&NPE\cite{NPE}&200MHZ&16-bit&0.208&10.40&1.00×&1.00×\\
~&VCK5000&CAT&\makecell{AIE:1.25GHZ\\PL 300MHZ}&INT8&\textbf{35.194}&\textbf{520.968}&\textbf{169.20×}&\textbf{50.09×}\\

\hline

\end{tabular}
\end{table*}

\begin{figure}[!t]
\centering
\includegraphics[width=3.5in]{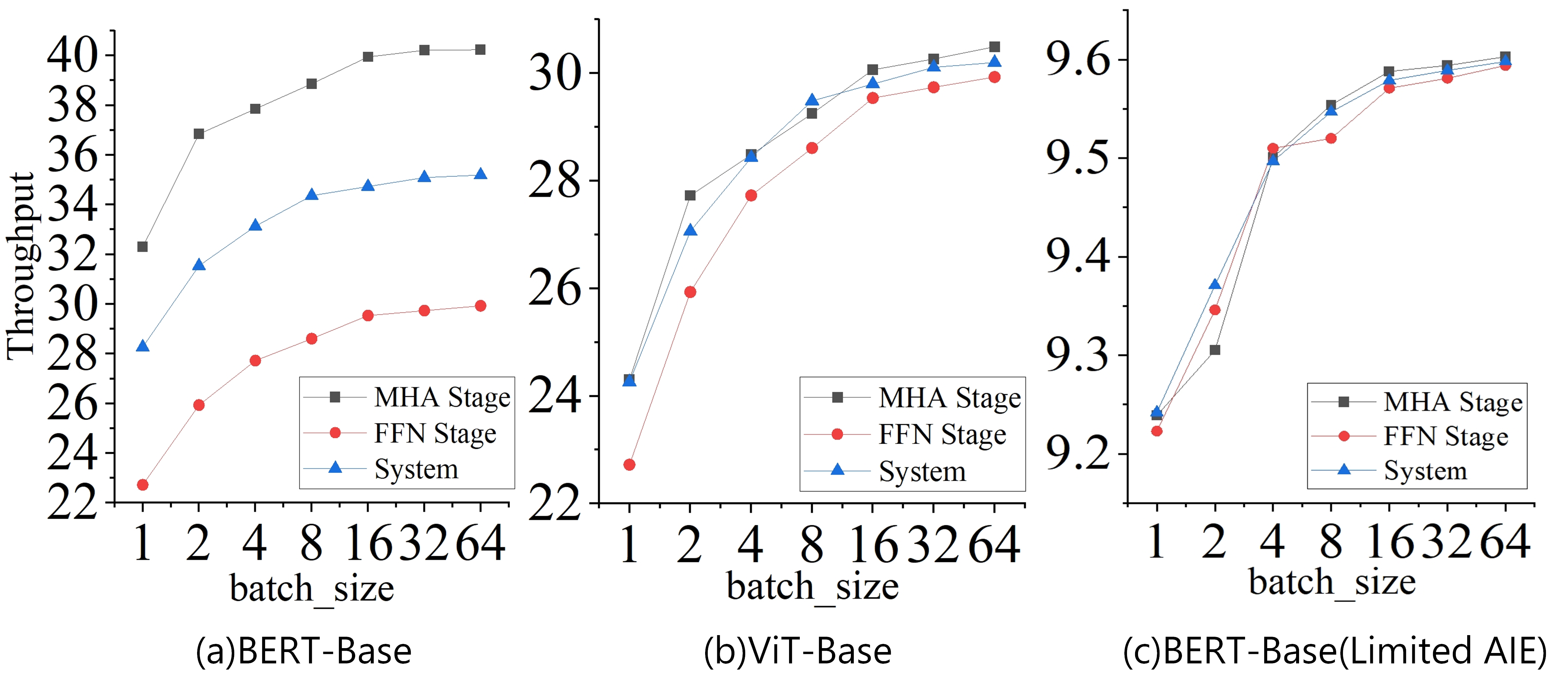}
\caption{Performance of three accelerators designed based on CAT framework under different batch\_size}
\label{fig_6}
\end{figure}

Figure 5 shows the performance performance of the three accelerators we designed based on the CAT framework under different batch\_size, and all the three accelerators tend to be stable at batch\_size of 16, which is close to the peak performance. Among them, although the throughput of BERT-Base and ViT-Base is reduced due to the pipeline start-up time in small batches, they still remain above 22TOPS, which is still at a high level. At the same time, it can also be seen in Figure 5 that the overall system performance is mostly between MHA Stage and FFN Stage, which also shows our AIE computing resource scheduling way more clearly.

\subsection{Performance Comparison}

Here, in order to further demonstrate the accelerator performance, we select six accelerators implemented on GPU, FPGA, and ACAP as comparison solutions. We compare the performance and energy efficiency of various Transformer accelerators, and then study these accelerators in depth by comparing and analyzing their peak performance and performance under specified models. Table VII lists the best performance for each accelerator and its detailed values for the specified model performance, respectively.

In Table VII, we first compare the peak throughput of each accelerator and the energy efficiency at peak throughput. SSR\cite{SSR} is the SOTA scheme before our CAT framework was proposed. Our accelerator has 1.31 times the peak throughput and 1.15 times the energy efficiency of SSR, which can surpass the current SOTA scheme. In addition, in traditional FPGA accelerator solutions such as (ViA\cite{FPGA_ACC_VIA}, Auto-ViT-Acc\cite{FPGA_ACC_AUTOVITACC}), we can achieve up to 113.9x throughput improvement and up to 64.78x energy efficiency improvement. On the NVIDIA A10G GPU accelerator\cite{GPU_ACC_A10G}, we were able to achieve 2.41x throughput improvement and 7.8x energy efficiency improvement.

For ViT models, for accelerators currently deployed on AMD Alveo U50, AMD ZCU102 and AMD VCK190 boards, accelerators deployed on the CAT framework achieve 97.9x, 42.6x and 1.37x throughput improvements on AMD VCK5000 boards. The energy efficiency is improved by 62.2 times, 5.85 times and 1.37 times.

For the BERT model, for the accelerator currently deployed on the Zynq Z-7100 board, the accelerator deployed on the CAT framework achieves 169.2 times throughput improvement and 50.09 times energy efficiency improvement on the AMD VCK5000 board.

\section{Conclusion and Future Work}

This paper introduces the CAT Framework, a customized Transformer accelerator deployment framework. CAT generates customized accelerator families through an abstract structure accelerator framework and the customization strategy of CAT framework. The CAT framework is able to exploit the best overall performance for Versal ACAP when deploying Transformer accelerators. Experimental results show that compared with the existing SOTA scheme, the accelerator customized based on the CAT framework can effectively accelerate various classical Transformer models.

In the future, we will explore extending/evaluating the CAT framework on larger models and more kinds of models. After that, the Transformer model can be further subdivided to build a more complete automatic deployment framework for AIE Transformer accelerators, which can promote the power of AIE processors in their advantage fields. Moreover, the Versal ACAP hardware still has room for optimization and can achieve even better performance if improved.

\bibliographystyle{IEEEtran}
\bibliography{sample-base}

\vspace{-100 mm}

 \begin{IEEEbiography}[{\includegraphics[width=1in,height=1.25in,clip,keepaspectratio]{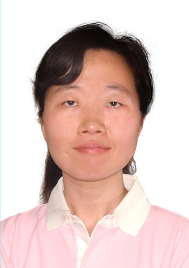}}]{Wenbo Zhang}
is an associate professor in FIT, Beijing University of Technology (BJUT), Beijing, China. Her research interests include heterogeneous computing, intelligent computing system and their applications.
\end{IEEEbiography}

\vspace{-100 mm}

 \begin{IEEEbiography}[{\includegraphics[width=1in,height=1.25in,clip,keepaspectratio]{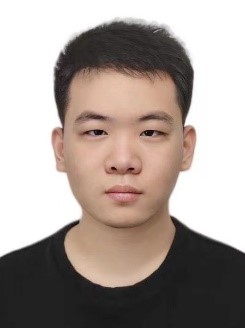}}]{Yiqi Liu}
is currently pursuing a master's degree in faculty of information technology (FIT), Beijing University of Technology (BJUT). His research interests include heterogeneous computing, intelligent computing system and their applications.
\end{IEEEbiography}

\vspace{-100 mm}

 \begin{IEEEbiography}[{\includegraphics[width=1in,height=1.25in,clip,keepaspectratio]{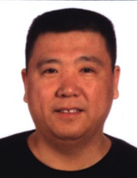}}]{Zhenshan Bao}
is an associate professor in faculty of information technology (FIT), Beijing University of Technology (BJUT), Beijing, China. His research interests include internet of things, intelligent computing system and their applications.
\end{IEEEbiography}

\end{document}